\documentclass{article} 
\usepackage[preprint]{arxiv-version}

\usepackage{microtype}
\usepackage{hyperref}
\usepackage{url}
\usepackage{booktabs}
\usepackage{wrapfig}
\usepackage{times}
\usepackage{latexsym}
\usepackage{listings}
\usepackage{adjustbox}
\usepackage{amsmath}
\usepackage{amssymb}
\usepackage{amsthm}
\usepackage{graphicx}
\usepackage{subcaption} 
\usepackage{float}
\usepackage{afterpage}
\usepackage[skip=2pt]{caption}
\usepackage{multirow}
\usepackage{booktabs}
\usepackage{float} 
\usepackage{caption}
\usepackage{cleveref}
\usepackage{tabularx} 
\usepackage{booktabs}  

\usepackage{array}     

\usepackage{CJKutf8}

\newcommand{\framework}[0]{WebProber}

\usepackage{graphicx}
\usepackage{subcaption}

\usepackage{lineno}

\definecolor{darkblue}{rgb}{0, 0, 0.5}
\hypersetup{colorlinks=true, citecolor=darkblue, linkcolor=darkblue, urlcolor=darkblue}

\title{AI Agents for Web Testing: A Case Study in the Wild}

\author{
\centerline{%
Naimeng Ye\thanks{Equal contribution.},\;\,%
Xiao Yu\footnotemark[1],\;\,%
Ruize Xu\footnotemark[1],\;\,%
Tianyi Peng,\;\,%
Zhou Yu} \vspace{5pt}
\\
\centerline{Columbia University} \vspace{5pt}\\
\centerline{\texttt{\{ny2336, xy2437, rx2246, tp2845, zy2461\}@columbia.edu}}
}

\begin{document}


\maketitle

\begin{abstract}
Automated web testing plays a critical role in ensuring high-quality user experiences and delivering business value. Traditional approaches primarily focus on code coverage and load testing, but often fall short of capturing complex user behaviors, leaving many usability issues undetected. The emergence of large language models (LLM) and AI agents opens new possibilities for web testing by enabling human-like interaction with websites and a general awareness of common usability problems. In this work, we present WebProber, a prototype AI agent-based web testing framework. Given a URL, WebProber autonomously explores the website, simulating real user interactions, identifying bugs and usability issues, and producing a human-readable report. We evaluate WebProber through a case study of 120 academic personal websites, where it uncovered 29 usability issues—many of which were missed by traditional tools. Our findings highlight agent-based testing as a promising direction while outlining directions for developing next-generation, user-centered testing frameworks.
\end{abstract}

\section{Introduction}
The modern web hosts billions of websites \citep{num-webs}, offering rich services and content that span nearly every aspect of daily life.
Common web applications include e-commerce websites such as Amazon, social media platforms like Facebook, information portals like Wikipedia along with a vast number of personal websites.

\begin{figure}[h]
    \centering
    \begin{minipage}[b]{0.48\textwidth}
        \centering
    \includegraphics[width=\textwidth]{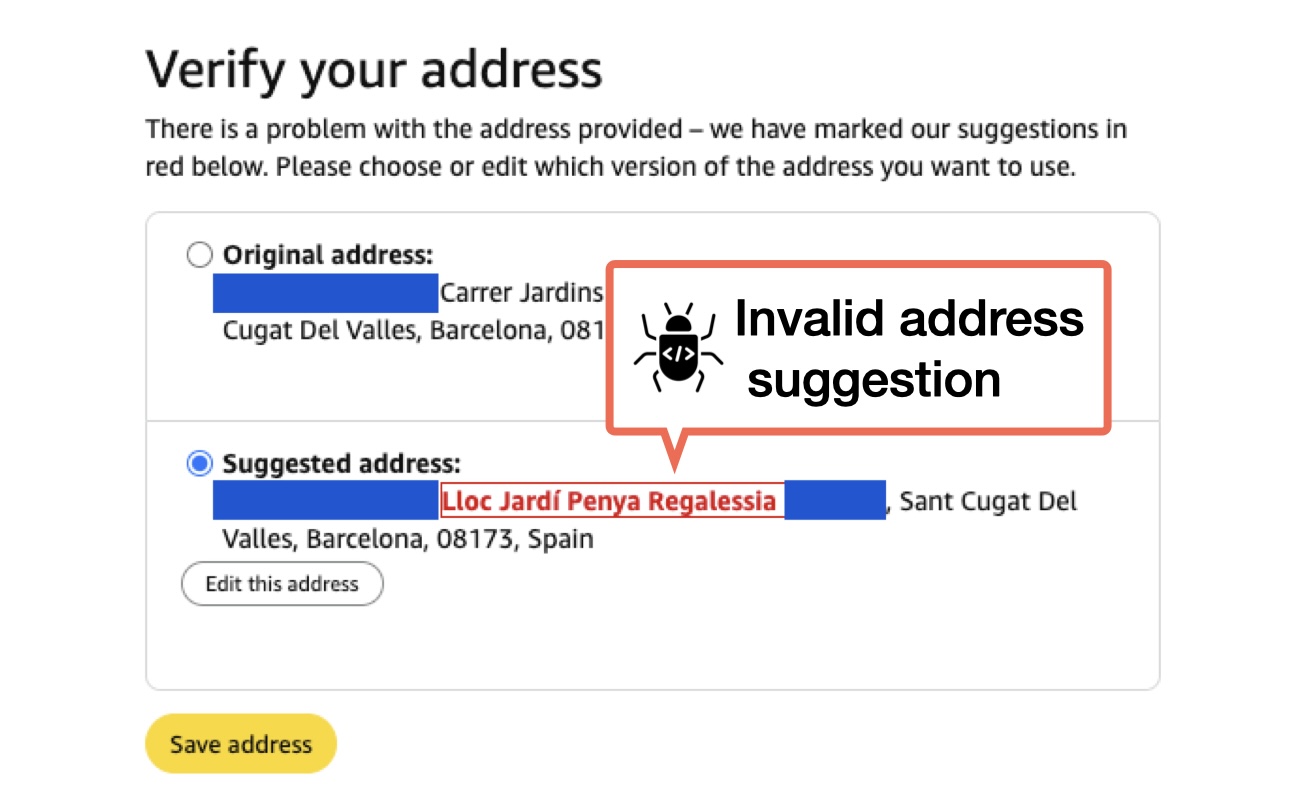}
        \subcaption{Amazon}
        \label{fig:sub1}
    \end{minipage}
    \hfill
    \begin{minipage}[b]{0.48\textwidth}
        \centering
        \includegraphics[width=\textwidth]{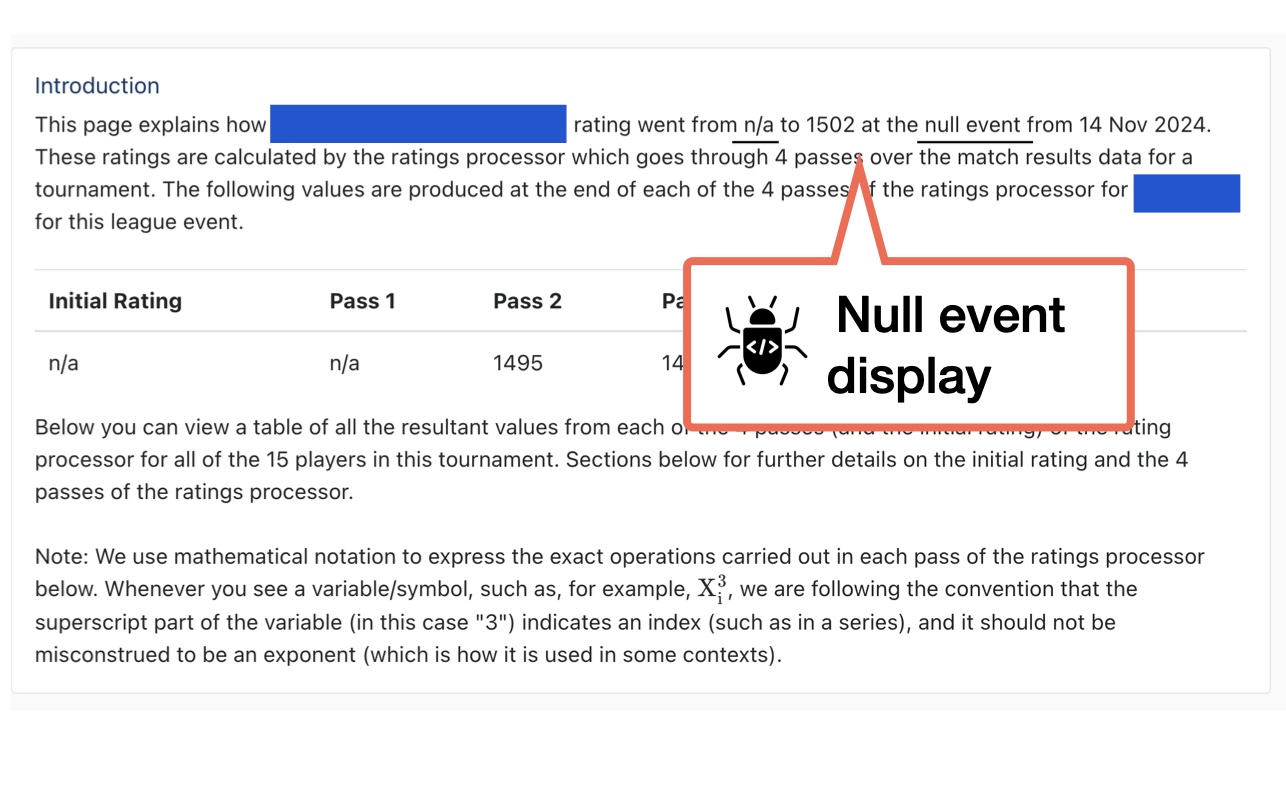}
        \subcaption{USA Table Tennis (USATT) Official Website}
        \label{fig:sub2}
    \end{minipage}

    \vspace{0.5cm}

    \begin{minipage}[b]{0.48\textwidth}
        \centering
        \includegraphics[width=\textwidth]{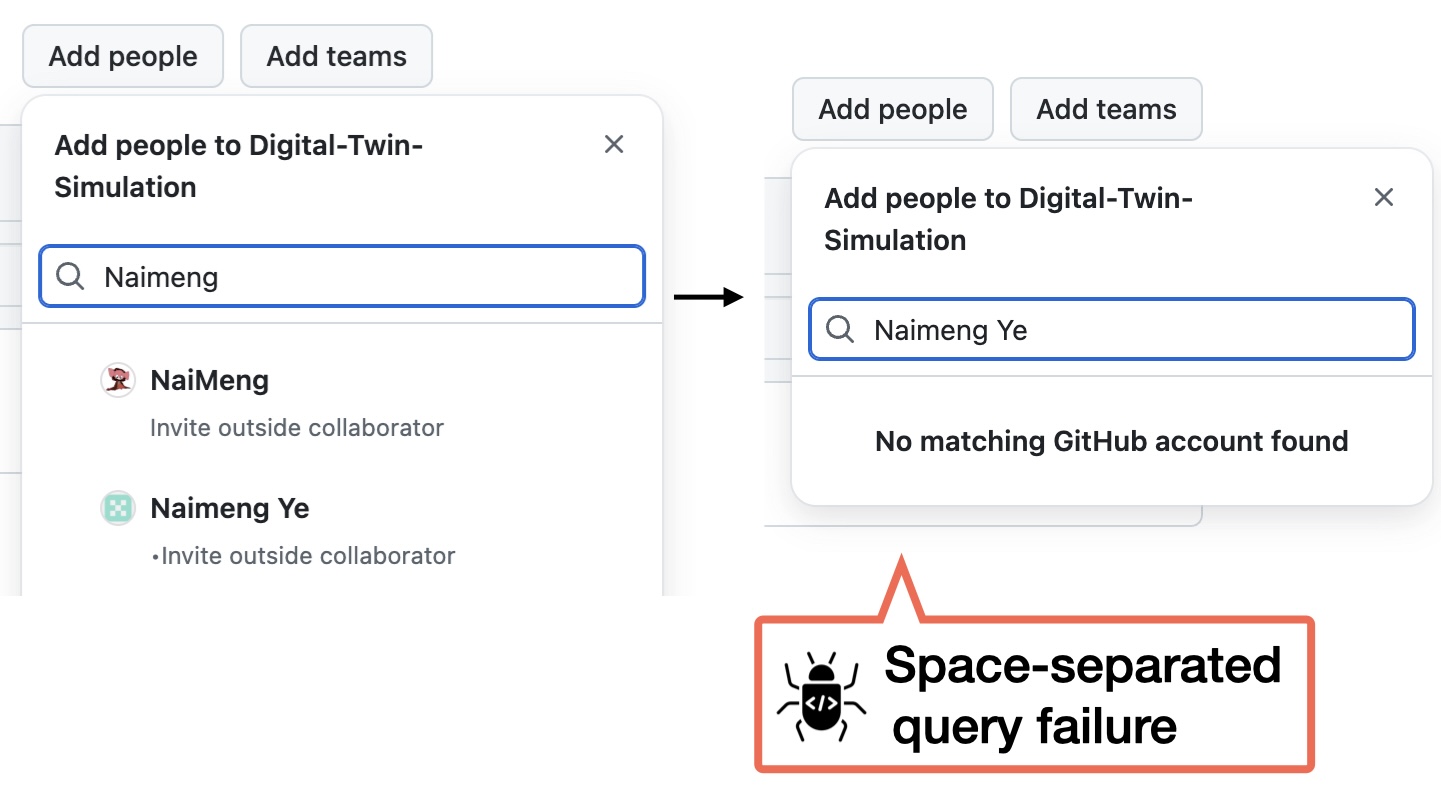}
        \subcaption{GitHub}
        \label{fig:sub3}
    \end{minipage}
    \hfill
    \begin{minipage}[b]
    {0.48\textwidth}
        \centering
        \includegraphics[width=\textwidth]{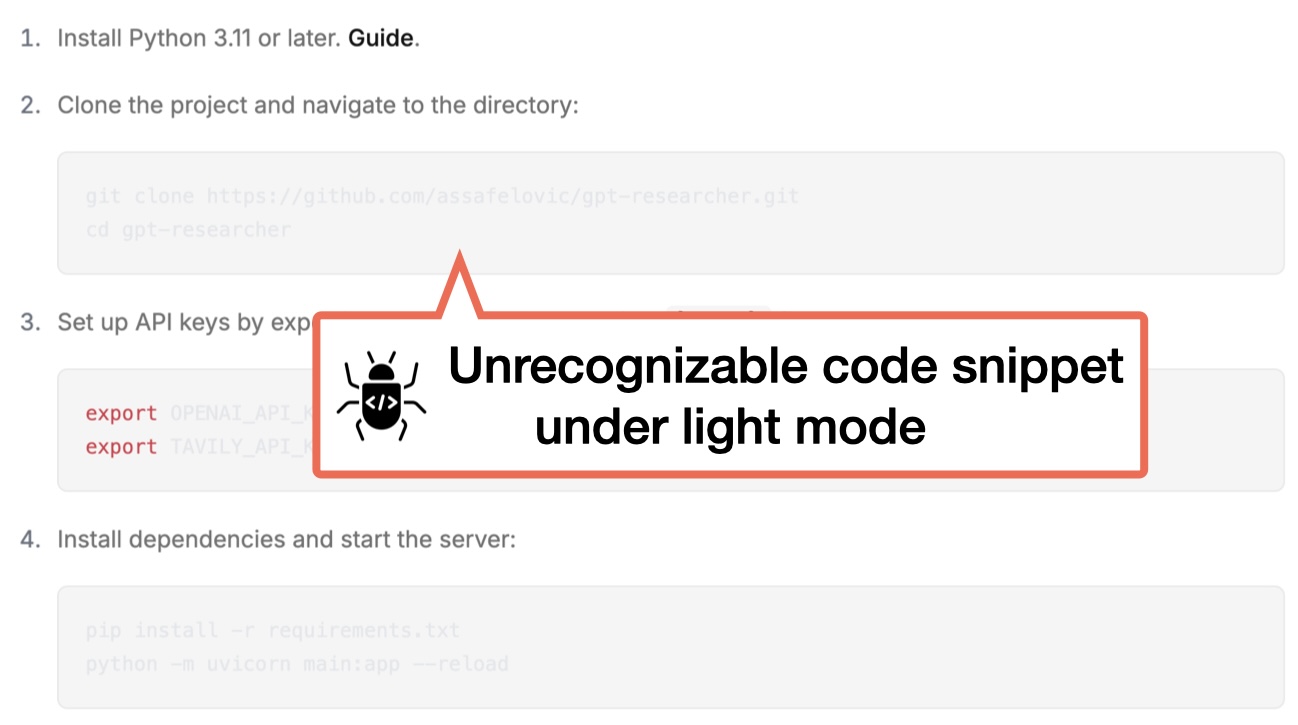}
        \subcaption{README Code Snippet}
        \label{fig:sub4}
    \end{minipage}
    \vspace{2em}
    \caption{Website usability bugs that are not easily detected by traditional web testing techniques. (a) On the Amazon Spain website, during a purchase, the system suggests a non-existent and unclickable address.
(b) The USATT website displays null event text for a league event.
(c) In a GitHub organization repository, the user search function does not support queries with spaces when adding users.
(d) On certain MCP server pages, code snippets in the README file are illegible in light mode due to poor color contrast.}
    \label{fig:intro-bug-examples}
\end{figure}

To ensure the quality and reliability of these web applications, automated web testing has become a critical component of modern web development cycles.
Traditional web testing approaches, such as static and dynamic analysis \citep{ricca2001analysis}, have been crucial in mitigating common issues and vulnerabilities such as layout and functional bugs.
These traditional approaches mainly rely on verifying code paths, automating scripted UI interactions, and measuring load performance using established tools like Cypress, Puppeteer, and JMeter \citep{cypress2025,puppeteer2025,jmeter2025}.
Despite these efforts, such approaches face significant challenges in detecting real-world, user-facing issues.
Since real users' actions are highly diverse and context-dependent, software-based methods often fail to cover test-cases that capture the full spectrum of user behavior.
This results in many undetected bugs and missing features that degrade user experience (see \Cref{fig:intro-bug-examples} for real-world examples).

\begin{figure}
    \centering
    \includegraphics[width=\linewidth]{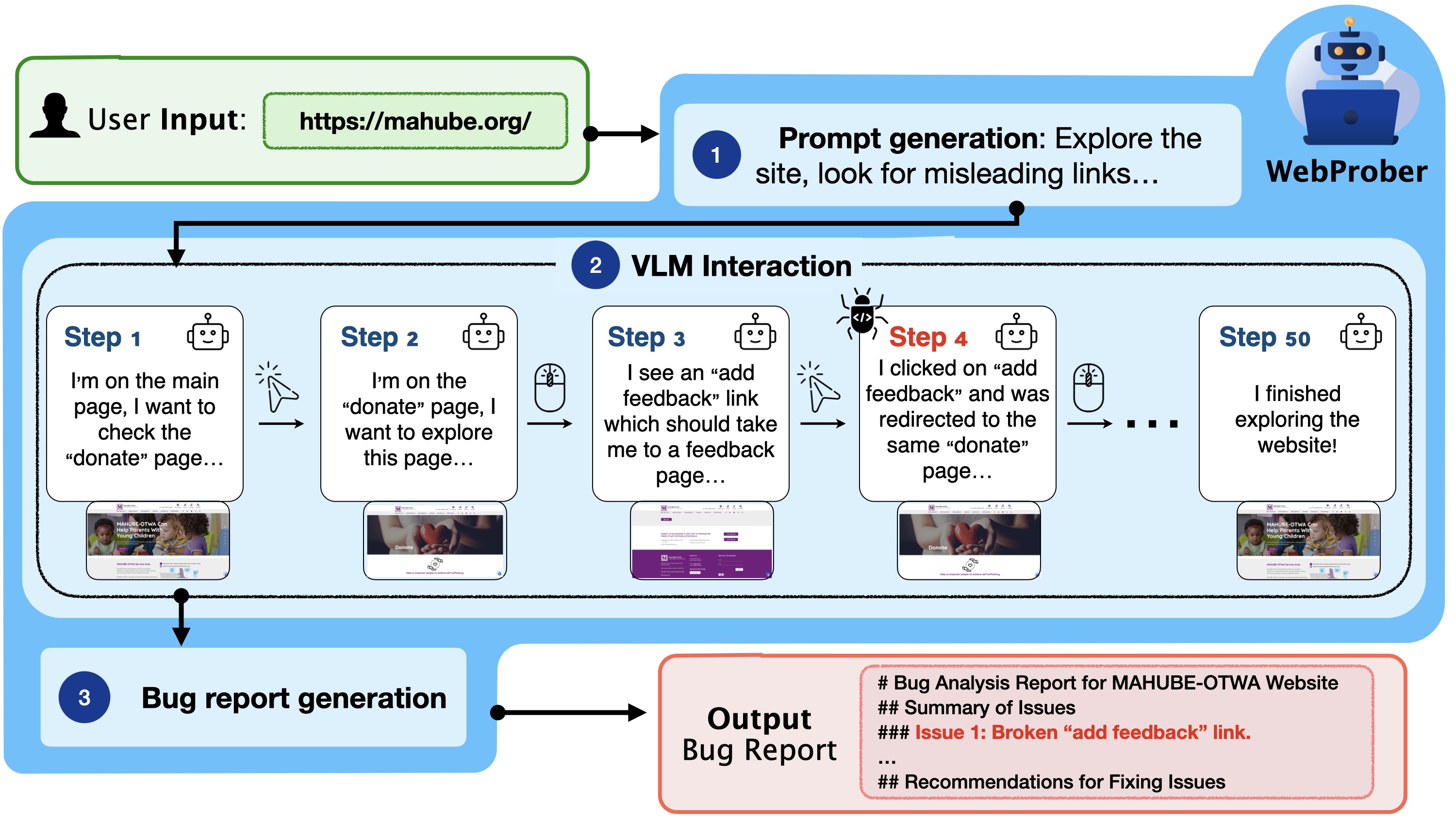}
    \caption{Workflow of \framework{}. Given a user-provided URL, the agent generates a comprehensive bug report through three stages: (1) testing prompt generation, (2) VLM-guided interaction, and (3) bug report generation. For more details, refer to \cref{sec:framework}.
    }
\label{fig:WebProbe-pipeline}
\end{figure}

We introduce \textbf{\framework{}}, a highly extensible web testing framework that leverages AI agents to simulate complex human behaviors on the web. Unlike existing approaches \citep{le2025automatedwebapplicationtesting,wang2025agentabautomatedscalableweb,lu2025uxagent} that use large language models to generate test cases or interact with post-processed HTML files, \framework{} employs powerful visual language models (VLMs) \citep{bordes2024introduction} to interact directly with visual webpages like human testers. Given a URL, \framework{} explores the webpage for common user-side bugs by performing actions such as clicking, typing, and scrolling. It generates a comprehensive report of unexpected website behaviors based on its interaction history.
We illustrate this workflow in \Cref{fig:WebProbe-pipeline}, which consists of three stages.
(1) a proposal module that suggests error-prone features to investigate, guided by a bug database, (2) an interaction module that simulates user experience guided by VLMs, and (3) a report generation module that examines the full interaction history to identify user-side bugs and suggest potential UI/UX improvements.

As a case study, we deployed our framework on 120 personal websites in the wild and found that our framework is able to identify 29 usability issues that impact user experience. Many of these issues---such as textual errors and misdirected links---were not detected by traditional automated testing tools, highlighting the unique strengths of agent-based testing in uncovering subtle, human-centric problems. 
Our empirical study on personal websites presents a first step towards building a scalable web testing framework based on AI agents, and we hope that this work can serve as a foundation for future research in this direction.

In summary, our contributions are:
\begin{enumerate}
    \item We introduce \framework{}, a highly extensible web testing framework that leverages AI agents to simulate human behavior on the web.
    \item We present a case study on 120 personal websites in the wild, on which \framework{} found 29 usability issues.
    \item We release our code and our human-annotated bug database for future research.\footnote{The github repository is available at \href{https://github.com/TianyiPeng/WebProber}{https://github.com/TianyiPeng/WebProber}. The database url is available at \href{https://webbugvid.netlify.app/}{https://webbugvid.netlify.app/}. Due to privacy concerns with some personal websites, we included only a subset of the bugs in the database.}
\end{enumerate}
\section{Related Work}
\paragraph{Browser-Use Agents}
With the advent of visual language models (VLMs), many works have explored the use of powerful models like GPT-4o and Claude-3.7 for web navigation tasks \citep{liu2023agentbenchevaluatingllmsagents,zhou2024webarenarealisticwebenvironment,koh2024vwa}.
Earlier efforts used the accessibility tree or a screenshot of the webpage as input to the VLM, and prompted it to generate actions such as clicking and typing \citep{yang2023setofmarkpromptingunleashesextraordinary,koh2024vwa}.
Recent works have explored various strategies to further improve an agent's decision-making process, such as iteratively prompting the model to improve its own output \citep{madaan2023selfrefineiterativerefinementselffeedback,shinn2023reflexionlanguageagentsverbal}, or augmenting the agent's decision process using search algorithms such as breadth- or depth-first search \citep{yao2023treethoughtsdeliberateproblem}, best-first search \citep{koh2024treesearchlanguagemodel}, and Monte Carlo tree search \citep{yu2023promptbasedmontecarlotreesearch,yu2025exactteachingaiagents}.
However, these works typically focus on solving pre-defined tasks such as finding a specific item on a shopping website, or navigating to a specific webpage.
Our work aims to use agents to \emph{discover} bugs missed by existing automated testing tools on real-world websites.

\paragraph{Automated Web Testing} Automated web testing emphasizes systematically testing web applications with minimal human intervention. Traditional approaches aim to generate action trajectories and can be broadly categorized into three classes: (1) randomized testing, where action sequences are generated stochastically \cite{androidMonkey2022}; (2) model-based methods, which construct a state graph of the application and use graph traversal algorithms such as depth-first search to explore it \cite{mesbah2012crawling,stocco2023neural,liu2025judge}; and (3) techniques based on reinforcement learning, which generate action sequences while maximizing a reward signal \cite{zheng2021automatic,sherin2023qexplore}.

More recently, the field has begun to incorporate LLMs in automated web testing. They are used to expand the test action space \cite{liu2023fill,liu2024testing,wang2024leveraging}, and to guide navigation and interaction 
\cite{alian2025feature,shahbandeh2024naviqate,liu2025temac}. In parallel, similar trends have emerged in mobile app testing \cite{liu2024make,yoon2023autonomous,lee2024mobilegpt,wen2024autodroid,chen2025standing}. Our work differs from prior literature by emphasizing the simulation of realistic user behaviors powered by VLMs, and by targeting contextual bugs often overlooked by traditional techniques, such as critical typographical errors or misdirected links.

\section{\framework{}}
\label{sec:framework}




We present \framework, a web testing system based on AI agents. Given a website URL, \framework{} returns a detailed report enumerating user-side bugs and UI/UX issues found during its interaction with the website.
\framework{} operates through a three-stage pipeline: (1) generating testing prompts that target common vulnerabilities for the particular class of website given, (2) simulating human-like web interactions, and (3) analyzing the interaction trajectory to generate comprehensive bug reports. We present an overview of this process in \Cref{fig:WebProbe-pipeline}, and describe this process in detail below.

\paragraph{Prompt generation} Prompts guide an AI agent to look for common usability issues for different classes of web pages, enabling more targeted and efficient exploration by focusing on typical usability issue patterns. In this work, we created our testing prompts through an iterative refinement process and release the final high-quality prompt template in our repository. For each website type (e.g., personal websites), we begin with a preliminary prompt instructing the VLM on which features to test and what issues to detect. We then refine both the prompt and bug set through iterative cycles: applying \framework{} to discover new bugs, manually verifying their reproducibility, and using a VLM to generate improved prompts based on the expanded bug set. This process continuously develops our prompt instructions while building a diverse collection of web usability bugs valuable for future evaluation. An example of prompt refinement is provided in \cref{subsec:proposing-potential-bugs}. While we demonstrate one effective approach to prompt generation, any method that produces high-quality, targeted prompts for usability testing would be compatible with our framework.

\paragraph{Interaction simulation} Using the generated testing prompts from the previous stage, \framework{} employs VLM-based agents to systematically interact with the website. Building on the Browser-Use Python package \citep{browser_use2024}, our system iteratively (1) prompts a VLM for an action based on a website screenshot (e.g., clicking a button or entering text), (2) executes the action on the website, and (3) repeats until either the maximum step limit is reached or the target feature has been tested. Throughout this process, we preserve the complete interaction trajectory, including screenshots, reasoning traces, and actions.

\paragraph{Bug report generation} Finally, we generate detailed bug and usability reports by analyzing the full interaction trajectory with a VLM. Since usability issues typically emerge during interactive use, the complete interaction history is important for an accurate diagnosis. Detailed prompts for report generation are provided in \cref{subsec:generating-bug-reports}.

In our implementation, we used Claude-3.7 Sonnet (\cite{claude-3.7}) as the VLM for each stage of \framework's pipeline. Each stage can be independently configured to use a different VLM, though exploring that is left as future work.

\section{Experiments}
To demonstrate the effectiveness of \framework{}, we conducted a case study on real-world academic personal websites crawled from OpenReview author profiles. We collected $120$ personal websites and applied \framework{} on this dataset to detect usability issues. We then manually inspected the generated reports to analyze the detected bugs, specifically evaluating whether they represented genuine usability issues or false positives. The results are presented in Section~\ref{sec:result}.

In addition to measuring the capabilities of \framework{}, we also investigated the coverage of bugs detectable by our framework. Since the total set of bugs on a website is unknown \emph{a priori}, we manually inspected a representative subset of $80$ websites to identify all potential bugs as a proxy for ground truth. We then ran \framework{} on the same subset of websites to investigate both detected and undetected issues. The results and analysis are presented in the following sections.

\subsection{Results}
\label{sec:result}
\begin{figure}[htbp]
    \centering
    \begin{subfigure}{0.49\linewidth}
        \centering
        \includegraphics[width=\linewidth]{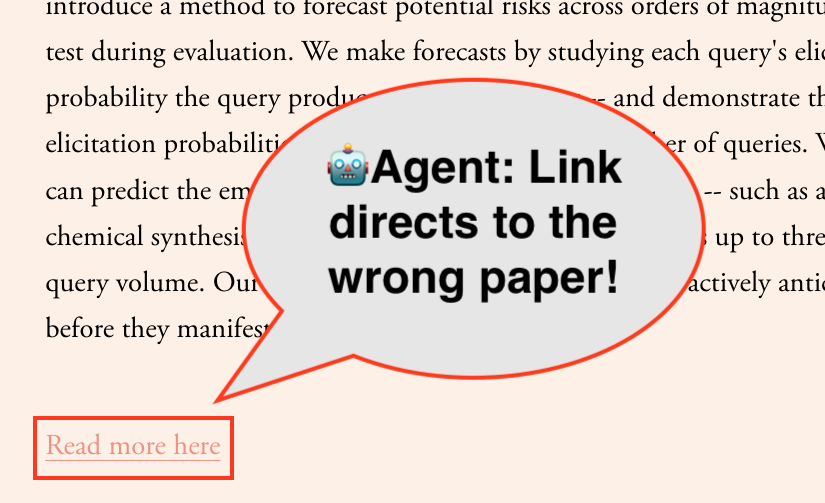}
        \caption{Misdirected link}
        \label{fig:wrong_link_bug}
    \end{subfigure}
    \hfill
    \begin{subfigure}{0.49\linewidth}
        \centering
        \includegraphics[width=\linewidth]{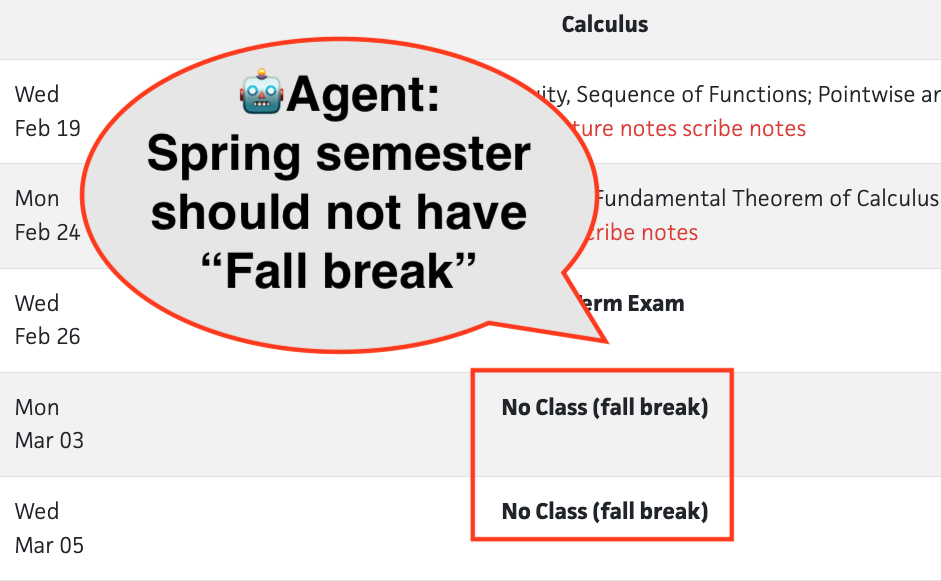}
        \caption{Typographical error}
        \label{fig:typo_bug}
    \end{subfigure}
    \vspace{5pt}
    \caption{Examples of \framework{} bug detection results. (a) A "Read more here" link for one research project incorrectly leads to a different paper. (b) A spring course syllabus mistakenly lists "fall break."}
    \label{fig:webprober_examples}
\end{figure}

\paragraph{Our approach effectively identifies usability issues that impact user experience}
Across our dataset of $120$ academic personal websites, \framework{} successfully identified $29$ usability issues (verified by the authors). In addition to bugs detectable by traditional techniques, e.g. rendering issues with images, our agent is also able to identify contextual bugs that are often overlooked by these methods. These issues span several categories, including link mistakes, rendering issues etc. We give a couple of representative examples of these usability issues as follows.
\begin{itemize}
    \item \textbf{Broken or misdirected links} The most common class of bugs detected is broken or misdirected links. We present an example in Figure~\ref{fig:wrong_link_bug}: the agent identified that a project description was inconsistent with the paper linked through the "Read more here" button.
    \item \textbf{Logical inconsistencies} Finally, we find our \framework{} is also able to detect logical inconsistencies in website contents, typically resulting from typographical errors. These errors sometimes lead to factual inaccuracies or user confusion. For example, in Figure~\ref{fig:typo_bug}, the agent identified a spring course syllabus (determined by calendar dates) that incorrectly scheduled a "Fall break" week.
\end{itemize}
These results illustrate the capabilities of VLM-based web testing and provide insights into what types of issue can be automatically detected.

\subsection{Discussion}
While \framework{} is able to identify real bugs and UI/UX issues in the wild, we also find numerous cases where human oversight is still needed for bug discovery. We present our findings below.

\begin{figure}
    \centering
    \includegraphics[width=\linewidth]{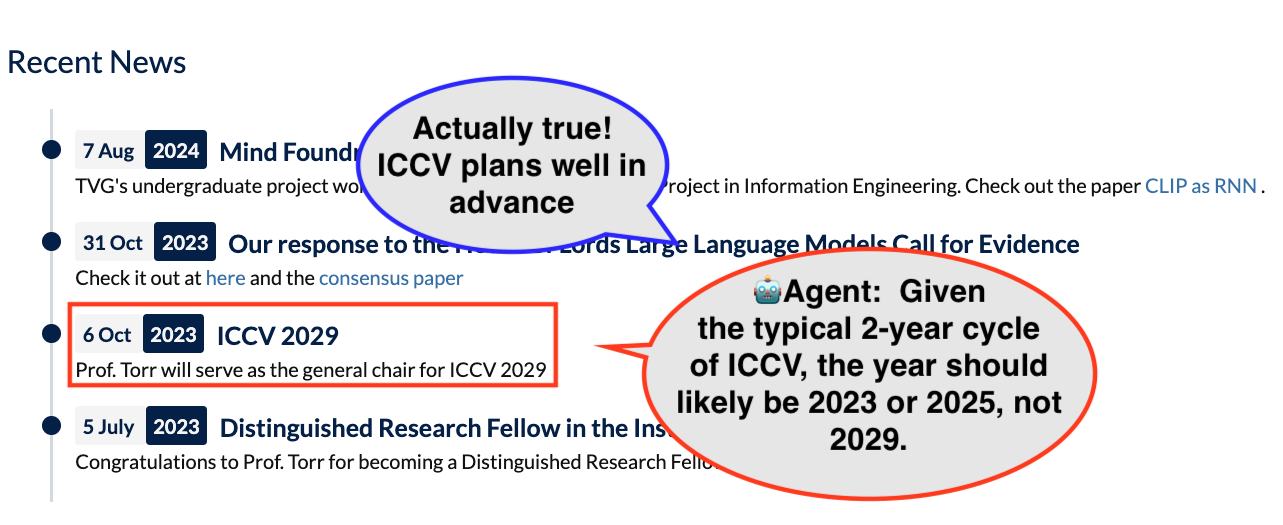}
    \caption{Example of false positive: an announcement referencing a ``2029'' conference is flagged as a typo, but in reality, conferences such as ICCV do plan leadership roles and organizational details several years in advance.}
    \label{fig:false_positive_typo}
\end{figure}

\paragraph{False positives}
While \framework{} successfully discovered $29$ usability issues, we found that $85\%$ of all reported bugs across the $120$ websites were false positives. The majority of these problems stemmed from technical limitations of the browser automation framework (the framework through which the agent applies actions on the webpage) rather than actual website issues. One common example is PDF access problems, which is often caused by the automation framework's security settings. However, the agent often incorrectly attributes these failures to website defects rather than automation constraints. Additionally, a small portion of false positives resulted from reasonable but incorrect assumptions of the agent, particularly when the agent lacked sufficient temporal or domain context to properly interpret website content. Figure~\ref{fig:false_positive_typo} illustrates one such example.
 
\paragraph{Undetected bugs}
On our representative subset of $80$ websites, we manually identified 32 bugs, of which \framework{} successfully detected $19$, achieving a coverage of $59.4\%$\footnote{
    Since the authors may not have found all possible bugs, the actual coverage may be lower.
}.
The undetected bugs fell in two primary failure modes. First, and most frequently, bugs were often located deep within the website hierarchy, requiring navigation through multiple pages. Since we executed our pipeline only once per website, the agent often terminated exploration before encountering these deeply embedded issues.
Second, certain pages containing bugs were inaccessible due to dynamic content rendering issues that our current implementation cannot handle effectively.
These results suggest that effective bug detection requires improved exploration strategies capable of performing systematic, long-horizon traversals of website hierarchies and handling dynamic content. We defer these enhancements to future work.

    
    
    

%

\section{Conclusion and Future Work}
We introduced \framework, an agent-based web testing framework. Applied to 120 academic personal websites, \framework{} uncovered 29 usability issues—many missed by traditional tools—demonstrating the potential of agent-driven testing. This case study also revealed several challenges and future directions:

\textbf{Agent-Browser Interaction.} Agent interactions remain unreliable—misclicks, erratic navigation, and poor performance on complex sites contribute to false positives. Enhancing browser control fidelity is a key priority.

\textbf{Bug Coverage and Training.} Current agents are not optimized for bug discovery. Reinforcement learning and hybrid approaches incorporating traditional automated web testing tools may improve coverage and effectiveness.

\textbf{Lack of Benchmarks.} Progress is hindered by the absence of a standardized benchmark for web usability issues. Curating datasets like SWEBench would support training and evaluation.

\textbf{Web Testing in Other Domains.} Vibe-coded websites, startup landing pages, and non-profit websites often involve quick prototyping with limited budgets for thorough quality assurance. AI-generated sites, in particular, may contain bugs that their creators—often without professional development expertise—are unable to detect.  While we believe AI agent-based web testing could significantly benefit these cases, we leave a rigorous field study for future work.

\bibliography{custom}

\begin{thebibliography}{38}
\providecommand{\natexlab}[1]{#1}
\providecommand{\url}[1]{\texttt{#1}}
\expandafter\ifx\csname urlstyle\endcsname\relax
  \providecommand{\doi}[1]{doi: #1}\else
  \providecommand{\doi}{doi: \begingroup \urlstyle{rm}\Url}\fi

\bibitem[Alian et~al.(2025)Alian, Nashid, Shahbandeh, Shabani, and
  Mesbah]{alian2025feature}
Parsa Alian, Noor Nashid, Mobina Shahbandeh, Taha Shabani, and Ali Mesbah.
\newblock Feature-driven end-to-end test generation.
\newblock In \emph{2025 IEEE/ACM 47th International Conference on Software
  Engineering (ICSE)}, pp.\  678--678. IEEE Computer Society, 2025.

\bibitem[{Android Developers}(2022)]{androidMonkey2022}
{Android Developers}.
\newblock {Monkey}.
\newblock \url{https://developer.android.com}, 2022.
\newblock Accessed: 2025-06-25.

\bibitem[Anthropic(2025)]{claude-3.7}
AI~Anthropic.
\newblock Claude 3.7 and claude code, 2025.
\newblock URL \url{https://www.anthropic.com/news/claude-3-7-sonnet}.

\bibitem[{Apache Software Foundation}(2025)]{jmeter2025}
{Apache Software Foundation}.
\newblock Apache jmeter: Load testing for web applications.
\newblock \url{https://jmeter.apache.org/}, 2025.
\newblock Accessed: 2025-06-26.

\bibitem[Bordes et~al.(2024)Bordes, Pang, Ajay, Li, Bardes, Petryk, Ma{\~n}as,
  Lin, Mahmoud, Jayaraman, et~al.]{bordes2024introduction}
Florian Bordes, Richard~Yuanzhe Pang, Anurag Ajay, Alexander~C Li, Adrien
  Bardes, Suzanne Petryk, Oscar Ma{\~n}as, Zhiqiu Lin, Anas Mahmoud, Bargav
  Jayaraman, et~al.
\newblock An introduction to vision-language modeling.
\newblock \emph{arXiv preprint arXiv:2405.17247}, 2024.

\bibitem[Chakarov(2023)]{num-webs}
Radoslave Chakarov.
\newblock How many websites are there? how many are active in 2023?
\newblock \url{https://webtribunal.net/blog/how-many-websites}, 2023.

\bibitem[Chen et~al.(2025)Chen, Liu, Chen, Wang, Wu, Hu, and
  Wang]{chen2025standing}
Mengzhuo Chen, Zhe Liu, Chunyang Chen, Junjie Wang, Boyu Wu, Jun Hu, and Qing
  Wang.
\newblock Standing on the shoulders of giants: Bug-aware automated gui testing
  via retrieval augmentation.
\newblock \emph{Proceedings of the ACM on Software Engineering}, 2\penalty0
  (FSE):\penalty0 825--846, 2025.

\bibitem[{Cypress.io}(2025)]{cypress2025}
{Cypress.io}.
\newblock Cypress: Testing frameworks for javascript.
\newblock \url{https://www.cypress.io/}, 2025.
\newblock Accessed: 2025-06-26.

\bibitem[{Google Chrome Developers}(2025)]{puppeteer2025}
{Google Chrome Developers}.
\newblock Puppeteer: Headless chrome node.js api.
\newblock \url{https://pptr.dev/}, 2025.
\newblock Accessed: 2025-06-26.

\bibitem[Koh et~al.(2024{\natexlab{a}})Koh, Lo, Jang, Duvvur, Lim, Huang,
  Neubig, Zhou, Salakhutdinov, and Fried]{koh2024vwa}
Jing~Yu Koh, Robert Lo, Lawrence Jang, Vikram Duvvur, Ming~Chong Lim, Po-Yu
  Huang, Graham Neubig, Shuyan Zhou, Ruslan Salakhutdinov, and Daniel Fried.
\newblock Visualwebarena: Evaluating multimodal agents on realistic visual web
  tasks, 2024{\natexlab{a}}.
\newblock URL \url{https://arxiv.org/abs/2401.13649}.

\bibitem[Koh et~al.(2024{\natexlab{b}})Koh, McAleer, Fried, and
  Salakhutdinov]{koh2024treesearchlanguagemodel}
Jing~Yu Koh, Stephen McAleer, Daniel Fried, and Ruslan Salakhutdinov.
\newblock Tree search for language model agents, 2024{\natexlab{b}}.
\newblock URL \url{https://arxiv.org/abs/2407.01476}.

\bibitem[Le et~al.(2025)Le, Bui, Nguyen, Nguyen, Vo, Luu, Nomura, and
  Nguyen]{le2025automatedwebapplicationtesting}
Nguyen-Khang Le, Quan~Minh Bui, Minh~Ngoc Nguyen, Hiep Nguyen, Trung Vo, Son~T.
  Luu, Shoshin Nomura, and Minh~Le Nguyen.
\newblock Automated web application testing: End-to-end test case generation
  with large language models and screen transition graphs, 2025.
\newblock URL \url{https://arxiv.org/abs/2506.02529}.

\bibitem[Lee et~al.(2024)Lee, Choi, Lee, Wasi, Choi, Ko, Oh, and
  Shin]{lee2024mobilegpt}
Sunjae Lee, Junyoung Choi, Jungjae Lee, Munim~Hasan Wasi, Hojun Choi, Steve Ko,
  Sangeun Oh, and Insik Shin.
\newblock Mobilegpt: Augmenting llm with human-like app memory for mobile task
  automation.
\newblock In \emph{Proceedings of the 30th Annual International Conference on
  Mobile Computing and Networking}, pp.\  1119--1133, 2024.

\bibitem[Liu et~al.(2025{\natexlab{a}})Liu, Gu, Wu, Zhang, Wei, and
  Xie]{liu2025temac}
Chenxu Liu, Zhiyu Gu, Guoquan Wu, Ying Zhang, Jun Wei, and Tao Xie.
\newblock Temac: Multi-agent collaboration for automated web gui testing.
\newblock \emph{arXiv preprint arXiv:2506.00520}, 2025{\natexlab{a}}.

\bibitem[Liu et~al.(2025{\natexlab{b}})Liu, Wang, Yang, Zhang, and
  Xie]{liu2025judge}
Chenxu Liu, Junheng Wang, Wei Yang, Ying Zhang, and Tao Xie.
\newblock Judge: Effective state abstraction for guiding automated web gui
  testing.
\newblock \emph{ACM Transactions on Software Engineering and Methodology},
  2025{\natexlab{b}}.

\bibitem[Liu et~al.(2023{\natexlab{a}})Liu, Yu, Zhang, Xu, Lei, Lai, Gu, Ding,
  Men, Yang, Zhang, Deng, Zeng, Du, Zhang, Shen, Zhang, Su, Sun, Huang, Dong,
  and Tang]{liu2023agentbenchevaluatingllmsagents}
Xiao Liu, Hao Yu, Hanchen Zhang, Yifan Xu, Xuanyu Lei, Hanyu Lai, Yu~Gu,
  Hangliang Ding, Kaiwen Men, Kejuan Yang, Shudan Zhang, Xiang Deng, Aohan
  Zeng, Zhengxiao Du, Chenhui Zhang, Sheng Shen, Tianjun Zhang, Yu~Su, Huan
  Sun, Minlie Huang, Yuxiao Dong, and Jie Tang.
\newblock Agentbench: Evaluating llms as agents, 2023{\natexlab{a}}.
\newblock URL \url{https://arxiv.org/abs/2308.03688}.

\bibitem[Liu et~al.(2023{\natexlab{b}})Liu, Chen, Wang, Che, Huang, Hu, and
  Wang]{liu2023fill}
Zhe Liu, Chunyang Chen, Junjie Wang, Xing Che, Yuekai Huang, Jun Hu, and Qing
  Wang.
\newblock Fill in the blank: Context-aware automated text input generation for
  mobile gui testing.
\newblock In \emph{2023 IEEE/ACM 45th International Conference on Software
  Engineering (ICSE)}, pp.\  1355--1367. IEEE, 2023{\natexlab{b}}.

\bibitem[Liu et~al.(2024{\natexlab{a}})Liu, Chen, Wang, Chen, Wu, Che, Wang,
  and Wang]{liu2024make}
Zhe Liu, Chunyang Chen, Junjie Wang, Mengzhuo Chen, Boyu Wu, Xing Che, Dandan
  Wang, and Qing Wang.
\newblock Make llm a testing expert: Bringing human-like interaction to mobile
  gui testing via functionality-aware decisions.
\newblock In \emph{Proceedings of the IEEE/ACM 46th International Conference on
  Software Engineering}, pp.\  1--13, 2024{\natexlab{a}}.

\bibitem[Liu et~al.(2024{\natexlab{b}})Liu, Chen, Wang, Chen, Wu, Tian, Huang,
  Hu, and Wang]{liu2024testing}
Zhe Liu, Chunyang Chen, Junjie Wang, Mengzhuo Chen, Boyu Wu, Zhilin Tian,
  Yuekai Huang, Jun Hu, and Qing Wang.
\newblock Testing the limits: Unusual text inputs generation for mobile app
  crash detection with large language model.
\newblock In \emph{Proceedings of the IEEE/ACM 46th international conference on
  software engineering}, pp.\  1--12, 2024{\natexlab{b}}.

\bibitem[Lu et~al.(2025)Lu, Yao, Gu, Huang, Wang, Li, Gesi, He, Li, and
  Wang]{lu2025uxagent}
Yuxuan Lu, Bingsheng Yao, Hansu Gu, Jing Huang, Zheshen~Jessie Wang, Yang Li,
  Jiri Gesi, Qi~He, Toby Jia-Jun Li, and Dakuo Wang.
\newblock Uxagent: An llm agent-based usability testing framework for web
  design.
\newblock In \emph{Proceedings of the Extended Abstracts of the CHI Conference
  on Human Factors in Computing Systems}, pp.\  1--12, 2025.

\bibitem[Madaan et~al.(2023)Madaan, Tandon, Gupta, Hallinan, Gao, Wiegreffe,
  Alon, Dziri, Prabhumoye, Yang, Gupta, Majumder, Hermann, Welleck,
  Yazdanbakhsh, and Clark]{madaan2023selfrefineiterativerefinementselffeedback}
Aman Madaan, Niket Tandon, Prakhar Gupta, Skyler Hallinan, Luyu Gao, Sarah
  Wiegreffe, Uri Alon, Nouha Dziri, Shrimai Prabhumoye, Yiming Yang, Shashank
  Gupta, Bodhisattwa~Prasad Majumder, Katherine Hermann, Sean Welleck, Amir
  Yazdanbakhsh, and Peter Clark.
\newblock Self-refine: Iterative refinement with self-feedback, 2023.
\newblock URL \url{https://arxiv.org/abs/2303.17651}.

\bibitem[Mesbah et~al.(2012)Mesbah, Van~Deursen, and
  Lenselink]{mesbah2012crawling}
Ali Mesbah, Arie Van~Deursen, and Stefan Lenselink.
\newblock Crawling ajax-based web applications through dynamic analysis of user
  interface state changes.
\newblock \emph{ACM Transactions on the Web (TWEB)}, 6\penalty0 (1):\penalty0
  1--30, 2012.

\bibitem[Müller \& Žunič(2024)Müller and Žunič]{browser_use2024}
Magnus Müller and Gregor Žunič.
\newblock Browser use: Enable ai to control your browser, 2024.
\newblock URL \url{https://github.com/browser-use/browser-use}.

\bibitem[Ricca \& Tonella(2001)Ricca and Tonella]{ricca2001analysis}
F.~Ricca and P.~Tonella.
\newblock Analysis and testing of web applications.
\newblock In \emph{Proceedings of the 23rd International Conference on Software
  Engineering. ICSE 2001}, pp.\  25--34, 2001.
\newblock \doi{10.1109/ICSE.2001.919078}.

\bibitem[Shahbandeh et~al.(2024)Shahbandeh, Alian, Nashid, and
  Mesbah]{shahbandeh2024naviqate}
Mobina Shahbandeh, Parsa Alian, Noor Nashid, and Ali Mesbah.
\newblock Naviqate: Functionality-guided web application navigation.
\newblock \emph{arXiv preprint arXiv:2409.10741}, 2024.

\bibitem[Sherin et~al.(2023)Sherin, Muqeet, Khan, and
  Iqbal]{sherin2023qexplore}
Salman Sherin, Asmar Muqeet, Muhammad~Uzair Khan, and Muhammad~Zohaib Iqbal.
\newblock Qexplore: An exploration strategy for dynamic web applications using
  guided search.
\newblock \emph{Journal of Systems and Software}, 195:\penalty0 111512, 2023.

\bibitem[Shinn et~al.(2023)Shinn, Cassano, Berman, Gopinath, Narasimhan, and
  Yao]{shinn2023reflexionlanguageagentsverbal}
Noah Shinn, Federico Cassano, Edward Berman, Ashwin Gopinath, Karthik
  Narasimhan, and Shunyu Yao.
\newblock Reflexion: Language agents with verbal reinforcement learning, 2023.
\newblock URL \url{https://arxiv.org/abs/2303.11366}.

\bibitem[Stocco et~al.(2023)Stocco, Willi, Starace, Biagiola, and
  Tonella]{stocco2023neural}
Andrea Stocco, Alexandra Willi, Luigi Libero~Lucio Starace, Matteo Biagiola,
  and Paolo Tonella.
\newblock Neural embeddings for web testing.
\newblock \emph{arXiv preprint arXiv:2306.07400}, 2023.

\bibitem[Wang et~al.(2025)Wang, Hsu, Lu, Gu, Cui, Xie, Headean, Yao,
  Veeragouni, Liu, Nag, and Wang]{wang2025agentabautomatedscalableweb}
Dakuo Wang, Ting-Yao Hsu, Yuxuan Lu, Hansu Gu, Limeng Cui, Yaochen Xie, William
  Headean, Bingsheng Yao, Akash Veeragouni, Jiapeng Liu, Sreyashi Nag, and
  Jessie Wang.
\newblock Agenta/b: Automated and scalable web a/btesting with interactive llm
  agents, 2025.
\newblock URL \url{https://arxiv.org/abs/2504.09723}.

\bibitem[Wang et~al.(2024)Wang, Wang, Fan, Li, and Liu]{wang2024leveraging}
Siyi Wang, Sinan Wang, Yujia Fan, Xiaolei Li, and Yepang Liu.
\newblock Leveraging large vision-language model for better automatic web gui
  testing.
\newblock In \emph{2024 IEEE International Conference on Software Maintenance
  and Evolution (ICSME)}, pp.\  125--137. IEEE, 2024.

\bibitem[Wen et~al.(2024)Wen, Tian, Pavlov, Du, Li, Chang, Zhao, Liu, Liu,
  Zhang, et~al.]{wen2024autodroid}
Hao Wen, Shizuo Tian, Borislav Pavlov, Wenjie Du, Yixuan Li, Ge~Chang, Shanhui
  Zhao, Jiacheng Liu, Yunxin Liu, Ya-Qin Zhang, et~al.
\newblock Autodroid-v2: Boosting slm-based gui agents via code generation.
\newblock \emph{arXiv preprint arXiv:2412.18116}, 2024.

\bibitem[Yang et~al.(2023)Yang, Zhang, Li, Zou, Li, and
  Gao]{yang2023setofmarkpromptingunleashesextraordinary}
Jianwei Yang, Hao Zhang, Feng Li, Xueyan Zou, Chunyuan Li, and Jianfeng Gao.
\newblock Set-of-mark prompting unleashes extraordinary visual grounding in
  gpt-4v, 2023.
\newblock URL \url{https://arxiv.org/abs/2310.11441}.

\bibitem[Yao et~al.(2023)Yao, Yu, Zhao, Shafran, Griffiths, Cao, and
  Narasimhan]{yao2023treethoughtsdeliberateproblem}
Shunyu Yao, Dian Yu, Jeffrey Zhao, Izhak Shafran, Thomas~L. Griffiths, Yuan
  Cao, and Karthik Narasimhan.
\newblock Tree of thoughts: Deliberate problem solving with large language
  models, 2023.
\newblock URL \url{https://arxiv.org/abs/2305.10601}.

\bibitem[Yoon et~al.(2023)Yoon, Feldt, and Yoo]{yoon2023autonomous}
Juyeon Yoon, Robert Feldt, and Shin Yoo.
\newblock Autonomous large language model agents enabling intent-driven mobile
  gui testing.
\newblock \emph{arXiv preprint arXiv:2311.08649}, 2023.

\bibitem[Yu et~al.(2023)Yu, Chen, and
  Yu]{yu2023promptbasedmontecarlotreesearch}
Xiao Yu, Maximillian Chen, and Zhou Yu.
\newblock Prompt-based monte-carlo tree search for goal-oriented dialogue
  policy planning, 2023.
\newblock URL \url{https://arxiv.org/abs/2305.13660}.

\bibitem[Yu et~al.(2025)Yu, Peng, Vajipey, Cheng, Galley, Gao, and
  Yu]{yu2025exactteachingaiagents}
Xiao Yu, Baolin Peng, Vineeth Vajipey, Hao Cheng, Michel Galley, Jianfeng Gao,
  and Zhou Yu.
\newblock Exact: Teaching ai agents to explore with reflective-mcts and
  exploratory learning, 2025.
\newblock URL \url{https://arxiv.org/abs/2410.02052}.

\bibitem[Zheng et~al.(2021)Zheng, Liu, Xie, Liu, Ma, Hao, and
  Liu]{zheng2021automatic}
Yan Zheng, Yi~Liu, Xiaofei Xie, Yepang Liu, Lei Ma, Jianye Hao, and Yang Liu.
\newblock Automatic web testing using curiosity-driven reinforcement learning.
\newblock In \emph{2021 IEEE/ACM 43rd International Conference on Software
  Engineering (ICSE)}, pp.\  423--435. IEEE, 2021.

\bibitem[Zhou et~al.(2024)Zhou, Xu, Zhu, Zhou, Lo, Sridhar, Cheng, Ou, Bisk,
  Fried, Alon, and Neubig]{zhou2024webarenarealisticwebenvironment}
Shuyan Zhou, Frank~F. Xu, Hao Zhu, Xuhui Zhou, Robert Lo, Abishek Sridhar,
  Xianyi Cheng, Tianyue Ou, Yonatan Bisk, Daniel Fried, Uri Alon, and Graham
  Neubig.
\newblock Webarena: A realistic web environment for building autonomous agents,
  2024.
\newblock URL \url{https://arxiv.org/abs/2307.13854}.

\end{thebibliography}
\bibliographystyle{colm-submission/colm2025_conference}

\clearpage
\appendix





\setcounter{table}{0}
\renewcommand{\thetable}{A\arabic{table}}
\setcounter{figure}{0}
\renewcommand{\thefigure}{A\arabic{figure}}

\section{Example bug report generated by \framework{}}

\begin{figure}[h]
    \centering
    
    \begin{minipage}[b]{0.45\textwidth}
        \centering
    \includegraphics[width=\textwidth]{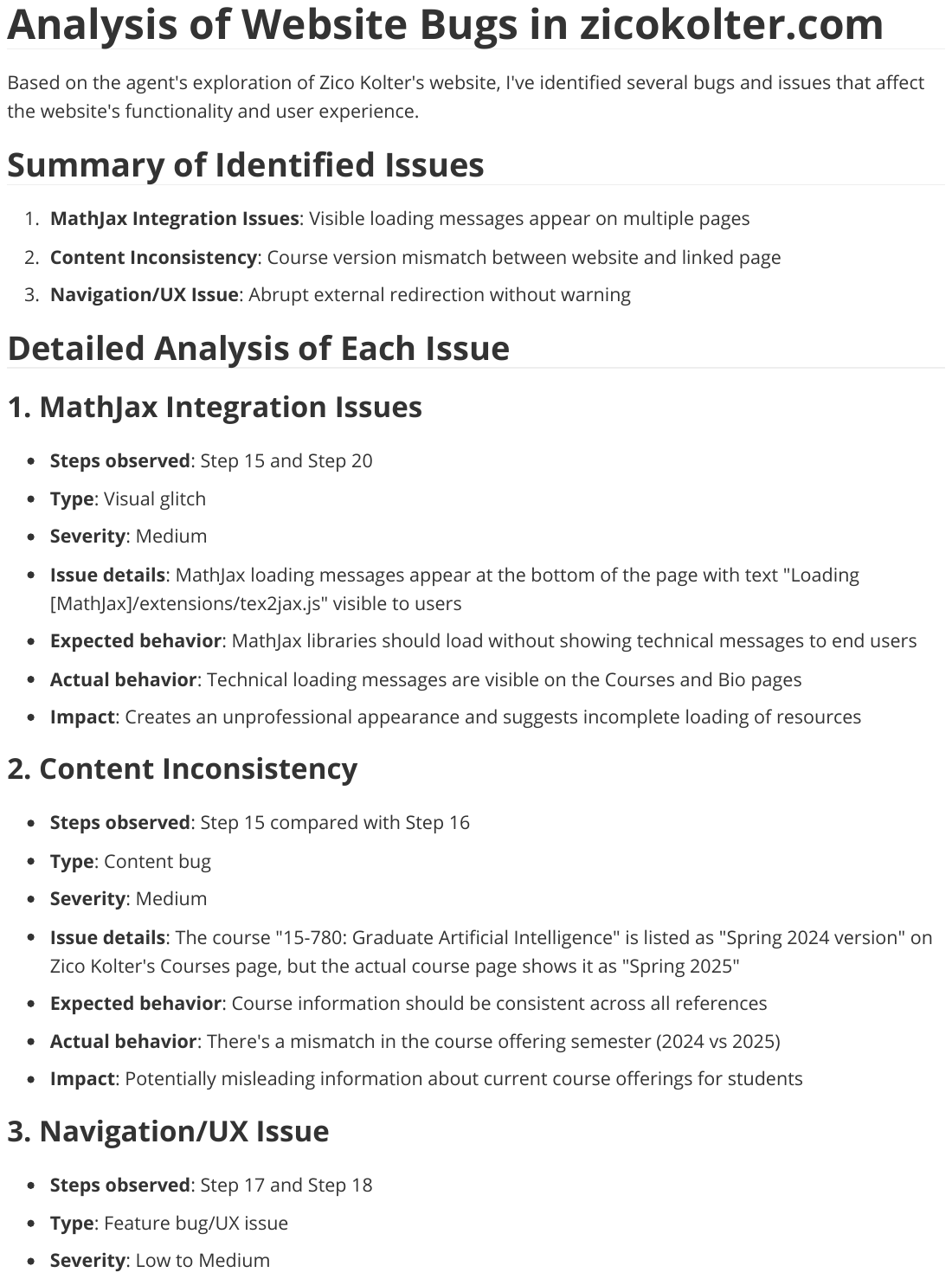}

    \end{minipage}
    \hfill
    \begin{minipage}[b]{0.45\textwidth}
        \centering
        \includegraphics[width=\textwidth]{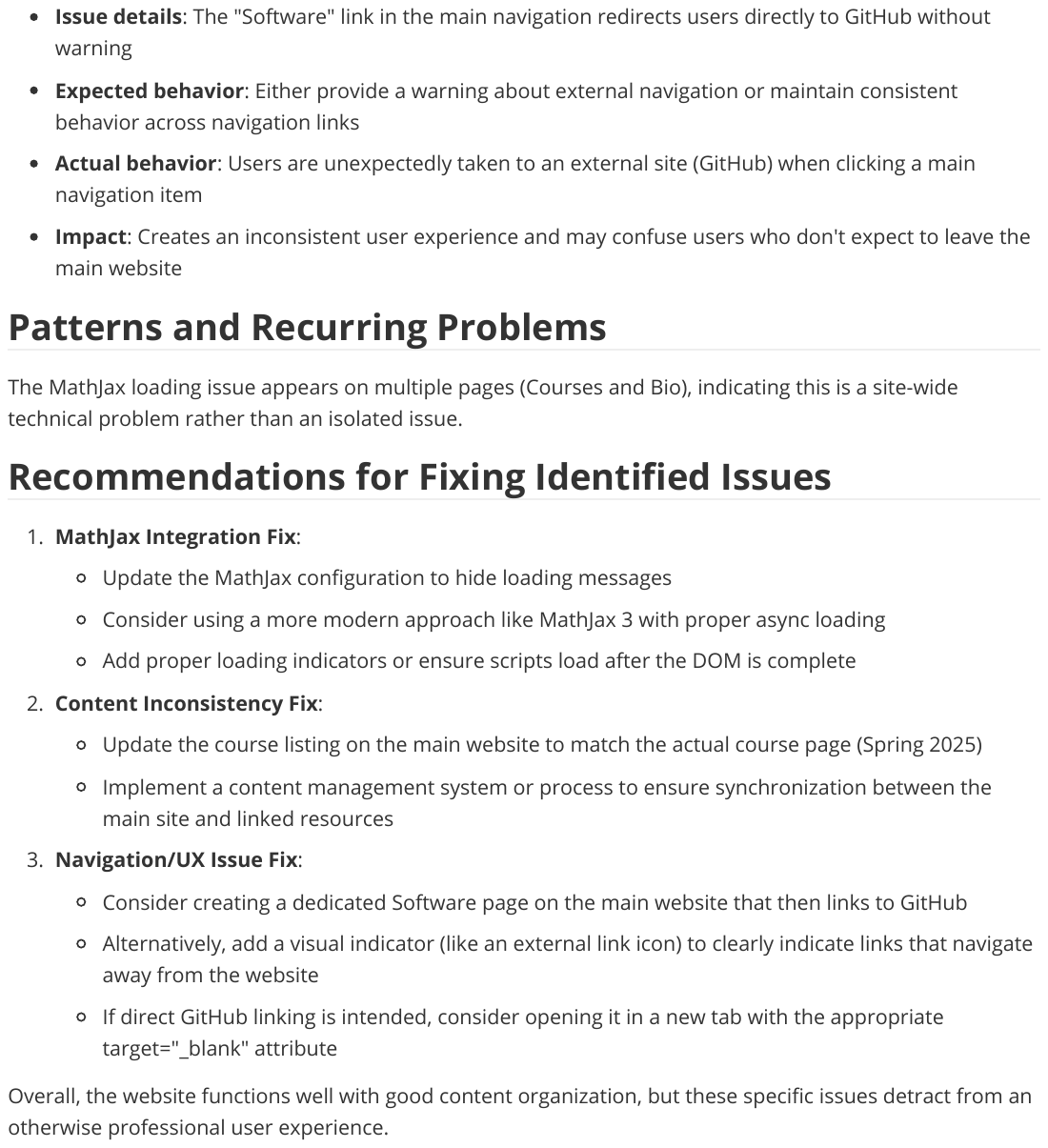}
       
    \end{minipage}

    \caption{An Example Bug Report on a Personal Website}
    
    \label{fig:zico-bug}
\end{figure}

Figure \ref{fig:zico-bug} shows an example of bug report by our \framework{}. Within 20 steps, the model correctly found the inconsistency between the dates of a course on the personal website and the course page. Via prompting, the report is well-formatted with bug descriptions and  details such as common patterns or recurring issues, and recommendations for fixing the bugs.

\newpage
\section{Implementation Details}

\subsection{Proposing Potential Bugs}
\label{subsec:proposing-potential-bugs}
To expand the database of historical issues, we follow an iterative prompting pipeline. Specifically, given the bugs found with current prompts, we pick up highly representative and domain-specific ones. Then, these bug descriptions are fed into powerful models to iteratively improve the current prompts. The updated prompts contain potential bug proposals that encourage the model to explore areas that are often more likely to be buggy. Table \ref{tbl:prompt} shows the initial prompt and the prompt iterated with ten historical bugs in the personal website domain.
\begin{table}[h]
    \caption{Initial and Iterated Prompts
    }
    \label{tbl:prompt}
    \begin{tabular}{p{1.52cm} p{11.51cm}}
        \toprule
        Iteration & Prompt\\
        \midrule
        
        Initial & Go to the website [URL], a personal website. Explore the content, click on links, and occasionally pause to assess whether what is shown and linked on the website is coherent and appropriate. Unreasonable or problematic issues include, but are not limited to: broken or mismatched images/links, UI glitches/incapabilities, illogical or unfunctional web design, or textual errors, etc.
        \\
        \cmidrule{2-2}
        One Iteration & 
        Go to the website [URL], a personal website. Systematically explore all accessible content, including menus, links, embedded media, interactive elements, and downloadable materials for WEBSITE BUGS. You should prioritize areas that are often more likely to contain issues. At each stage, critically evaluate whether the displayed information, layout, and behavior align with expectations for a functional and professional web experience. Carefully inspect for issues such as, but not limited to: 
        \newline
        
        (1) Broken elements: dead/missing links, 404 pages, failed image or video loads. 
        
        (2) Interaction failures: non-responsive buttons, malfunctioning forms or filters, non-working download or redirect actions. 
        
        (3) UI/UX flaws: lack of visual feedback, missing tooltips/ESC buttons, layout inconsistencies, uncustomized templates, poor mobile compatibility. 
        
        (4) Content inconsistencies: outdated or contradictory data (e.g., dates or names), mismatched references or external links, typos or formatting errors. 
        
        (5) Domain-specific bugs: for instance, broken external links to publications, projects, GitHub, Google Scholar, etc. Incorrect anchor links (e.g., internal navigation like \#about or \#projects not working). Outdated or dead email links (e.g., mailto: pointing to deprecated addresses). Missing or malformed citation info (e.g., BibTeX files, DOI links not rendering or downloading properly). Mismatched thumbnails or missing alt-text on research project previews. Videos or talks not embedded properly (e.g., iframe blocked by CORS).
\newline

       For each identified issue, consider its impact, repeatability,
and specific trigger (e.g., ”clicking X under condition Y leads to error Z”)\\
        \bottomrule
    \end{tabular}
\end{table}

\subsection{Generating Bug Reports}
\label{subsec:generating-bug-reports}
Given the trajectories, we then prompt powerful models such as Claude-3.7 Sonnet to analyze the trajectories and generate a formatted bug report. We prompt the model to contain a summary of bugs, the steps where bugs occurred, short descriptions, and then more details such as common patterns or recurring issues, and recommendations for fixing the bugs. The full prompt is shown in Table \ref{tbl:prompt_bug}.

\begin{table}[h]
    \caption{Prompt for Generating Bug Reports
    }
    \label{tbl:prompt_bug}
    \begin{tabular}{p{13.5cm}}
        \toprule
         Prompt\\
        \midrule
        Please analyze the following agent run trajectory and identify any potential bugs or glitches in the website being tested. Consider both feature bugs (missing or incorrect functionality) and glitch-like bugs (visual or behavioral anomalies). Note that the type of bug is not always obvious, so don't be afraid to make an assumption. For example, if the website does not support certain features that the agent is trying to use, that is a bug (e.g. the agent is trying to use the "add to cart" feature, but the website does not have a cart, or that the agent is searching in some language that the website does not support).
\newline

For each step, I'll provide:

0. The screenshot of the current browser state

1. The agent's evaluation of the step

2. The next goal

3. The action taken
\newline

Please analyze the entire sequence of steps and identify:

1. Any unexpected behaviors or errors of the website itself (*note: not the agent's actions*)

2. Missing or incorrect functionality

3. Visual glitches or UI inconsistencies

4. Any other anomalies that might indicate bugs
\newline

Here's the step-by-step trajectory:
\newline

[Trajectory]
\newline 

Based on the above trajectory, please provide:

1. A summary of any bugs or glitches identified

2. The specific steps where issues occurred

3. The nature of each issue (feature bug, visual glitch, etc.)

4. Any patterns or recurring problems

5. Recommendations for fixing the identified issues
\newline

For each identified issue, please specify:

- The step number where it occurred

- Whether it's a feature bug or visual glitch

- The severity of the issue

- The expected behavior vs actual behavior \\
        \bottomrule
        \end{tabular}
        
\end{table}

\end{document}